\documentclass[superscriptaddress,showpacs,nofootinbib,hyperref,tightenlines,twocolumn]{revtex4-1}
\usepackage{amsmath,verbatim,latexsym,amssymb,graphicx,subcaption,indentfirst,mathrsfs,amsthm,dsfont,hyperref}

\usepackage{natbib,url,textcase,bm}
\newcounter{tlc}
\newtheorem{theorem}[tlc]{Theorem}

\newtheoremstyle{indented}{5pt}{3pt}{\addtolength{\leftskip}{3.5em}}{}{\bfseries}{.}{.5em}{}
\theoremstyle{indented}

\makeatletter
\newcommand\footnoteref[1]{\protected@xdef\@thefnmark{\ref{#1}}\@footnotemark}
\makeatother

\usepackage{soul}

\usepackage{color}
\usepackage[dvipsnames]{xcolor}
\usepackage[normalem]{ulem}

\newcommand{\Rel}{D_1}

\newcommand{\Tr}{{\rm Tr}}

\newcommand{\QSys}{\mathcal{S}}
\newcommand{\Sites}{N}
\newcommand{\Min}{{\rm min}}
\newcommand{\Max}{{\rm max}}
\newcommand{\LParen}{\bm{(}}
\newcommand{\RParen}{\bm{)}}

\newcommand*{\bra}[1]{\langle #1\rvert}
\newcommand*{\ket}[1]{\lvert #1 \rangle}
\newcommand*{\braket}[2]{\langle #1 \lvert #2 \rangle}
\newcommand*{\ketbra}[2]{\lvert #1 \rangle\!\langle #2 \rvert}

\newcommand{\kB}{k_\mathrm{B}}
\newcommand{\fwd}{\mathrm{fwd}}
\newcommand{\rev}{\mathrm{rev}}
\newcommand{\diss}{\mathrm{diss}}
\newcommand{\forfeit}{\mathrm{forfeit}}
\newcommand{\worst}{\mathrm{worst}}
\DeclareMathOperator{\supp}{supp}

\begin{document}

\title{Maximum one-shot dissipated work from R\'{e}nyi divergences}

%
% Authors
%
\author{Nicole~Yunger~Halpern\footnote{E-mail: nicoleyh.11@gmail.com}}
\affiliation{Institute for Quantum Information and Matter, Caltech, Pasadena, CA 91125, USA}

\author{Andrew~J.~P.~Garner}
\affiliation{Atomic and Laser Physics, Clarendon Laboratory,
University of Oxford, Parks Road, Oxford OX1 3PU, United Kingdom}
\affiliation{Center for Quantum Technologies, National University of Singapore, Republic of Singapore}

\author{Oscar~C.~O.~Dahlsten\footnote{Email: dahlsten@sustc.edu.cn}}
\affiliation{Institute for Quantum Science and Engineering, and Department of Physics, Southern University of Science and Technology, Shenzhen 518055, China}
\affiliation{Atomic and Laser Physics, Clarendon Laboratory,
University of Oxford, Parks Road, Oxford OX1 3PU, United Kingdom}
\affiliation{London Institute for Mathematical Sciences,
35a South Street, Mayfair, London, W1K 2XF, United Kingdom}

\author{Vlatko~Vedral}
\affiliation{Atomic and Laser Physics, Clarendon Laboratory,
University of Oxford, Parks Road, Oxford OX1 3PU, United Kingdom}
\affiliation{Center for Quantum Technologies, National University of Singapore, Republic of Singapore}
\affiliation{Department of Physics, National University of Singapore, 2 Science Drive 3, Singapore 117542}

\date{\today}

%%
%% ABSTRACT
%%
\begin{abstract}
Thermodynamics describes large-scale, slowly evolving systems.
Two modern approaches generalize thermodynamics:
{\em fluctuation theorems}, which concern finite-time nonequilibrium processes, and {\em one-shot statistical mechanics}, which concerns small scales and finite numbers of trials. 
Combining these approaches, we calculate a one-shot analog of the average dissipated work defined in fluctuation contexts: the cost of performing a protocol in finite time instead of quasistatically.
The average dissipated work has been shown to be proportional to a relative entropy between phase-space densities, one between quantum states, and one between probability distributions over possible values of work.% ~\cite{GomezPB}. 
We derive one-shot analogs of all three equations, demonstrating that the order-infinity R\'{e}nyi divergence is proportional to the maximum possible dissipated work in each case.
These one-shot analogs of fluctuation-theorem results contribute to the unification of these two toolkits for small-scale, nonequilibrium statistical physics.
\end{abstract}

\maketitle{}

%
% Body
%

%
%
% Introduction
%
%
\section{Introduction}

Thermodynamics concerns large scales and infinitesimally slow evolutions. In the thermodynamic limit, a system's size approaches infinity and is typified by mean behaviors. \emph{Quasistatic} processes proceed slowly enough that 
the system remains in equilibrium.
Equilibrium quantities describe quasistatic processes---for example, the temperature $T$ and the free energy $F$ such as Helmholtz's, $- \kB T \ln Z$ (wherein $\kB$ denotes Boltzmann's constant and $Z$ denotes a partition function).

Two recently developed frameworks generalize thermodynamic concepts, such as work and heat, beyond slow processes and infinite sizes.
{\em Fluctuation relations} interrelate equilibrium quantities such as $F$ and nonequilibrium processes~(e.g.,~\cite{Jarzynski97,Crooks99,Tasaki00,KawaiPV07,QuanD08,Seifert12}). 
One-shot statistical mechanics is used to quantify 
the efficiency with which 
work can be invested or extracted, 
including outside the assumptions of conventional statistical mechanics
(e.g.,~\cite{DahlstenRRV11,EgloffDRV12,Aberg13,Dahlsten13,HorodeckiO13}):
First, the system may be small, violating the thermodynamic limit.
Second, the work performed in any given trial---rather 
than just the work averaged over trials---may be reasoned about.
Third, the system may occupy a quantum state
coherent relative to the energy eigenbasis.

One-shot statistical mechanics relies on the mathematical toolkit of 
{\em one-shot information theory}, or
{\em information theory beyond i.i.d.}
(independent and identically distributed variables and quantum states)
(e.g.,~\cite{Renyi61,Renner05,Datta09,DupuisKFRR12}).
Conventional information theory concerns information-processing tasks
such as data compression.
One assumes that $n$ random variables $X$,
or quantum states $\rho$, are processed.
The variables and states are assumed to be i.i.d.
For example, the probability $p_x$ that $X$ evaluates to $x$
is the same for all instances of $X$.
One calculates the optimal efficiency with which
the task can be performed,
on average over $n$,
in the limit as $n \to \infty$.
\emph{Asymptotic} entropies,
such as the Shannon and von Neumann entropies~\cite{NielsenC10},
quantify these efficiencies.
These entropies are generalized in one-shot information theory.
Examples include the R\'enyi divergences $D_\alpha$, discussed below.
The generalized entropies quantify the efficiencies with which
more-general information-processing tasks can be performed.
For example, few copies of $X$ or $\rho$ may be processed.
The variables or states may not be i.i.d. 

One-shot information theory generalizes
conventional information theory,
as one-shot statistical mechanics extends
conventional statistical mechanics.
A combination of fluctuation relations and one-shot statistical mechanics describes quite general thermodynamic systems~\cite{YungerHalpernGDV14}.

Transforming one equilibrium state quasistatically into another requires an amount $W$ of work equal to the difference between the states' free energies: $W = \Delta F$. Implementing a protocol in finite time yields a nonequilibrium state and costs extra work, some dissipated as heat. This penalty of irreversibility is called the {\em dissipated work}, or \emph{irreversible work}. The average $\langle W_{\rm diss} \rangle :=  \langle W \rangle  -  \Delta F$ over many trials has been studied in fluctuation contexts (e.g.~\cite{Crooks98,DeffnerL10,DornerGCPV12}).\footnote{
% < f >
Our discussion of work can be phrased alternatively in terms of entropy production (e.g.,~\cite{DeffnerL10}).}
% < /f >
We define the {\em one-shot dissipated work}  $W_\diss:=W-\Delta F$ as the penalty paid in one trial. 

$\langle W_\diss \rangle$ has been shown to be proportional to three instances of the \emph{Kullback-Leibler divergence}, or \emph{average relative entropy}, $\Rel$. 
$\Rel$ quantifies how much two probability distributions, or two quantum states, differ.
(See the ``R\'enyi divergences'' section and~\cite{NielsenC10} for reviews.)
$\langle W_\diss \rangle$ has been related to three average relative entropies:
(i) a $\Rel$ between phase-space densities 
$\rho(p,q,t)$ and $\tilde{\rho}(p,-q,t)$,
associated with forward and time-reversed processes~\cite{KawaiPV07};
(ii) a $\Rel$ between quantum states $\rho(t)$ and $\tilde{\rho}(t)$,
associated with forward and reverse processes~\cite{ParrondoVdBK09};
and 
(iii) a $\Rel$ between probability distributions $P_\fwd(W)$ and $P_\rev(-W)$
over the work performed during the forward and reverse processes.

This $\Rel$ belongs to a family of \emph{R\'enyi divergences} $D_\alpha$
that quantify the discrepancies between distributions or between states.
$\Rel$ quantifies a discrepancy in terms of an average over many copies of a distribution or state.
The {\em order-$\infty$ R\'enyi divergence} $D_\infty$ quantifies 
 the distinguishability apparent, in a worst case, from just one copy. 

We derive one-shot analogs of all three thermodynamic equalities.
The averages $\langle W_\diss \rangle$ and $\Rel$ are replaced with 
the one-shot $W_\diss^\worst$ and $D_\infty$.
The trio reveals the generality of the proportionality between 
the worst-case dissipated work and a one-shot entropy.

We begin by reviewing fluctuation theorems and R\'enyi divergences, 
focusing on $D_\infty$. We recall each 
$\langle W_\diss \rangle$ proportionality and derive its one-shot analog. 
Our main results relate the maximum possible penalty $W_{\diss}^{\rm worst}$ of investing work in finite time to three instances of $D_\infty$.
We apply our results to a quantum quench, 
 whose work distribution has been studied in several settings~\cite{Silva_08_Statistics,Gambassi_11_Statistics,Abeling_16_Quantum,DornerGCPV12,Lobejko_17_Work,Sindona_14_Statistics,Garcia_16_Quantum}.
Our one-shot analogs of fluctuation-relation results illustrate 
the insights offered by merging fluctuation relations with one-shot statistical mechanics.

\section{Background}

We review fluctuation theorems, then R\'{e}nyi divergences.

\subsection{Fluctuation theorems}
Consider a system governed by a time-dependent Hamiltonian $H (\lambda_t)$. The external parameter $\lambda_t$ changes in time: $t \in [-\tau, \tau]$. 
Suppose the system begins in the thermal state 
$\gamma_{-\tau}:=e^{-\beta H(\lambda_{-\tau})}/Z_{-\tau}$, wherein $\beta$ denotes a heat bath's  inverse temperature and $Z_{-\tau}$ normalizes the state. 
Suppose an agent switches $\lambda_t$ from $\lambda_{-\tau}$ to $\lambda_\tau$ while the system interacts with the bath. The switching costs work, the amount of which varies from trial to trial. 
A probability distribution $P_\fwd(W)$ represents the probability that a given trial costs work $W$. By $P_\rev(-W)$, we denote the probability that initializing the Hamiltonian to $H (\lambda_\tau)$ and initializing the system in $\gamma_{\tau}:=e^{-\beta H(\lambda_{\tau})}/Z_\tau$, then reversing the drive according to $\lambda_{-t}$, outputs work $W$. 

Fluctuation relations such as Crooks' Theorem govern these distributions~\cite{Crooks98}. Let $\Delta F := F(\gamma_\tau)-F(\gamma_{-\tau})$ denote the difference between the free energy of $\gamma_\tau$ and that of $\gamma_{-\tau}$. 
(Throughout this article, we shall assume $\Delta F$ is finite.)
Assuming the system is classical; coupled to a bath; and undergoing a Markovian, microscopically reversible evolution, Crooks proved that
\begin{equation}
\label{eq:Crooks}
\frac{P_\fwd(W)}{P_{\rm rev}(-W)}=e^{\beta(W-\Delta F)}
\end{equation}
\cite{Crooks98}. 
Identical theorems have been shown to govern quantum systems isolated from~\cite{Tasaki00}, or interacting with the bath while work is performed  (e.g.,~\cite{QuanD08}).

%
%
% Intro to divergences
%
%
\subsection{R\'{e}nyi divergences}

Let $P$ and $Q$ denote probability distributions over the set of values $\{x\}$.
%
%Let $X$ denote a random variable that can assume values in $\{ x \}$.
%Let $P$ and $Q$ denote probability distributions over the set.}
The \emph{order-$\alpha$ R\'{e}nyi divergence} quantifies 
the distinctness of $P$ and $Q$~\cite{Renyi61,vanErvenH12},
\begin{align}
   D_\alpha( P || Q )  :=
   \frac{1}{ \alpha - 1}   \ln  \left(   
   \int  dx  \;  P^\alpha(x)   Q^{1 - \alpha}(x)  \right),
\end{align}
or of quantum states $\rho$ and $\sigma$~\cite{Beigi13}: 
\begin{align}
   D_\alpha( \rho || \sigma )  :=
   \frac{1}{ \alpha - 1}   \ln  \LParen   
   \Tr ( \rho^\alpha  \sigma^{1 - \alpha} )   \RParen,
\end{align}
wherein Tr denotes the trace, for $\alpha \in [0, 1) \cup (1, \infty)$.

The order-$1$ R\'{e}nyi divergence, known also as the \emph{Kullback-Leibler divergence} and the \emph{average relative entropy}, follows from the limit as $\alpha\to1$:
\begin{equation}
   \Rel ( P || Q )
   =   \int dx  \;   P(x) \ln \LParen P(x)  / Q(x)\RParen
\end{equation}
for classical distributions, and
$\Rel(\rho||\sigma)={\rm Tr}\LParen\rho[\ln(\rho)-\ln(\sigma)]\RParen$ for quantum states.
$\Rel$ quantifies an average of the information learned
when one mistakes $Q$ for $P$, or $\sigma$ for $\rho$, then is corrected~\cite{CoverT06,Hiai_91_Proper}.

We focus on the order-$\infty$ divergences:
For classical distributions,
\begin{equation}
\label{eq:D_infty_def_class}
D_\infty(    P   ||   Q  )
=   \ln   \LParen
     {\rm min}  \{   \lambda \in \mathds{R}   \,  :  \, 
                            P(x)  \leq   \lambda   Q(x)   \; \: \forall x  \}   \RParen
\end{equation}
if the support $\supp(Q)\subseteq\supp(P)$,  
and $D_\infty(P||Q)=\infty$ otherwise.
For quantum states,
\begin{equation}
\label{eq:DefDinftyQuantum}
D_\infty(    \rho   ||   \sigma  )
   =   \ln   \left(   \max_{i,j}  \left\{
   \frac{r_i}{s_j}    \,   :    \,    \braket{r_i}{s_j}   \neq0    \right\}   \right)
\end{equation}
for quantum states $\rho=\sum_i r_i\ketbra{r_i}{r_i}$  and 
\mbox{$\sigma=\sum_j s_j\ketbra{s_j}{s_j}$}~\cite{TomamichelBH14}.
Imagine receiving just one copy of a state that is $\rho$ or $\sigma$.
Suppose, for simplicity, that the states share the eigenbasis
 $\{\ketbra{r_i}{r_i}\}$, which you measure.
In the worst case, two events occur:
 (1) The outcome, $i_0$, maximizes the ratio $r_{i_0}/s_{i_0}$.
Since $r_{i_0}$ is enormous, while $s_{i_0}$ is tiny, you guess that you received $\rho$.
(2) You then learn that you received $\sigma$.
% NYH: "Subsequently" is a good word; don't get me wrong! "Then" is much shorter, though.
% AJPG: Not shorter if the metric is by word ;)  Still, simpler is better.
% NYH: Yes, fewer syllables are easier for the reader to digest, regardless of what word count PRL's technology spew out.
The information gained from event (2), after (1), 
equals $D_\infty(\rho||\sigma)$.

%
%
% D_infty between phase-space densities
%
%
\section{Results}

We now derive equalities between the worst-case work and
(i) phase-space densities, (ii) quantum states, and 
(iii) work distributions.

\subsection{Divergences between phase-space densities}

Kawai {\em et al.} consider a classical system that remains isolated from the bath while work is performed~\cite{KawaiPV07}. Governed by Hamiltonian dynamics, the system follows a deterministic trajectory through phase space. Specifying a phase-space point $(q, p)$ at any time $t$ uniquely specifies a trajectory and a work cost $W(q, p, t)$.

An experimenter does not know which trajectory the system follows in any given forward trial, because the experimenter  ascribes to the system the initial state $e^{-\beta H( \lambda_{-\tau} ) } / Z_{-\tau}$. The probability that the system occupies an area-$(dq \: dp)$ region centered on $(q, p)$ at time $t$ is 
$\rho(q, p, t) \: dq \: dp$, wherein $\rho(q, p, t)$ denotes the phase-space density.
$\tilde{\rho}(q, p, t)$ denotes the phase-space density after an amount $\tilde{t} = 2\tau-t$ of time has passed during the reverse protocol.

Kawai \emph{et al.}\ proceed as follows.
As the system loses no heat while work is performed, the work required to evolve the system along some trajectory equals the difference between the final and initial Hamiltonians: $W(p,q,t)  =  H(q_{\tau},p_{\tau}, \tau) - H(q_{-\tau}, p_{-\tau},-\tau)$. 
The forward process's initial $\rho$ and the reverse process's initial $\tilde{\rho}$ are equated with thermal states. The Hamiltonian is assumed to have time-reversal invariance (TRI): $H(q, p, t) = H(q, -p, t)$. 
% Reference: Kawai '07, p. 2, LHS, first paragraph
From TRI, the preservation of phase-space densities by Hamiltonian dynamics, and the correspondence of $\rho(q, p, t)$ and $\tilde{\rho}(q, -p, t)$ to the same Hamiltonian follows the ``generalized Crooks relation''
\begin{equation}   \label{eq:genCrooks}
   e^{ \beta  [  W  \!\left(q, p,t\right) - \Delta F  ] }
    =
    \dfrac{\rho\left(q,p,t\right)}{\tilde{\rho}\left(q,-p,t\right)}.
\end{equation}
By taking logs, multiplying each side by $\tilde{\rho}(q, -p, t)$, and integrating over phase space, Kawai \emph{et al.}\ derive
\begin{align}    \label{eq:Kawai}
   \langle W_{\rm diss} \rangle
   =   \frac{1}{\beta}  D \LParen \rho(q, p, t)  ||  \tilde{\rho}(q, -p, t)  \RParen.
\end{align}
The right-hand side (RHS) is well-defined if the support of $\rho$ lies in the support of 
$\tilde{\rho}$:  
${\rm supp}\LParen\rho(q,p,t)\RParen\subseteq{\rm supp}\LParen\tilde{\rho}(q,-p,t)\RParen$~\cite{ParrondoVdBK09}.
% Reference: Parrondo, p. 5, near the top

The nonnegativity of $\Rel$ implies that, on average, performing a protocol quickly dissipates positive work. The work penalty's nonnegativity has been interpreted as the Second Law of Thermodynamics~\cite{KawaiPV07,Jarzynski08}. 
According to Stein's Lemma, $\Rel(P || Q)$ quantifies the average probability that an attempt to distinguish between $P$ and $Q$ will fail~\cite{CoverT06,VaikuntanathanJ09}.
$\Rel\LParen\rho(q,p,t)||\tilde{\rho}(q,-p,t)\RParen$ quantifies the distinguishability of the forward-process density from its time-reverse. 
$\Rel(P||Q)$ vanishes if and only if $P=Q$~\cite{CoverT06}.
Equation~\eqref{eq:RelS2} shows that reversing the trajectory followed during the forward protocol yields the trajectory followed during the reverse protocol if and only if the system dissipates no work on average. No work is dissipated if the process proceeds quasistatically, such that the system remains in equilibrium. Hence $\Rel$ quantifies roughly how far from equilibrium the system evolves.

Let us turn from averages over infinitely many trials to single trials, starting with our first theorem.

\begin{theorem}  \label{theorem:Kawai}
The worst-case dissipated work of the foregoing protocol is proportional to an order-$\infty$ R\'{e}nyi divergence between phase-space distributions:
\begin{align}  
   \label{eq:KawaiOne1}
   W_\diss^{\rm worst}
   =  \frac{1}{\beta}
   D_\infty\LParen\rho(q, p, t)  ||  \tilde{\rho} (q, -p, t)\RParen,
\end{align}
if ${\rm supp} \LParen\rho(q,p,t)\RParen\subseteq{\rm supp}\LParen\tilde{\rho}(q, -p, t)\RParen$.

\begin{proof}
First, we take the logarithm of each side of the generalized Crooks relation [Eq.~\eqref{eq:genCrooks}]:
\begin{align}  \label{eq:KawaiHelp1}
   W  -  \Delta F  =  \frac{1}{\beta}  
   \ln  \left(  \frac{  \rho(q, p, t) }{  \tilde{\rho} (q,  -p,  t) }  \right).
\end{align}
We maximize each side of the equation, invoking the logarithm's monotonicity to shift the maximum into the argument:
\begin{align} \label{eq:KawaiHelp2}
   W_{\rm max}  -  \Delta F  =  \frac{1}{\beta} 
   \ln  \left(  \max  \left\{   \frac{  \rho(q, p, t) }{  \tilde{\rho} (q,  -p,  t) }  \right\}  \right).
\end{align}
Comparing the left-hand side (LHS) with the definition of $W_\diss^{\rm worst}$ and the RHS with the definition of $D_\infty$ yields Eq.~\eqref{eq:KawaiOne1}.
\end{proof}
\end{theorem}

%
% Analysis of Kawai one-shotification
%
Like Eq.~\eqref{eq:Kawai}, Theorem~\ref{theorem:Kawai} relates dissipated work to a measure of the difference between $\rho(p, q, t)$ and $\tilde{\rho}(p, -q, t)$. 
The more work is dissipated during the most expensive possible trial, the less the forward-process density can resemble its time-reversed cousin, 
as measured by $D_\infty$. 
The lesser the resemblance, the farther the system is expected to depart from equilibrium.  
As in Eq.~\eqref{eq:Kawai}, the LHS of Eq.~\eqref{eq:KawaiOne1} is time-independent, so the RHS remains constant for all $t \in [-\tau, \tau]$.

Equation~\eqref{eq:KawaiOne1} has the correct quasistatic limit: If work is invested infinitesimally slowly, the worst amount of work that can be dissipated---the only amount that can be dissipated---vanishes: $W_{\rm max} - \Delta F = \Delta F - \Delta F = 0$. Because the system remains in equilibrium, $H (\lambda_t)$ and $\beta$ determine the state completely. The RHS of Ineq.~\eqref{eq:KawaiOne1} becomes 
$D_\infty\LParen\rho(q, p, t) || \tilde{\rho}(q, -p, t)\RParen=0$.

Theorem~\ref{theorem:Kawai} can aid an agent who has imperfect information about phase-space densities. Kawai \emph{et al.}\ recommend using Eq.~\eqref{eq:Kawai} to predict $\langle W_{\rm diss} \rangle$ from $\rho$ and $\tilde{\rho}$. Phase-space densities, they acknowledge, can be difficult to learn about. So they bound $\langle W_{\rm diss} \rangle$ with a $\Rel$ between  coarse-grained densities. Theorem~\ref{theorem:Kawai} offers an alternative to coarse-graining. One can use the theorem upon learning just the maximum of $\rho / \tilde{\rho}$, rather than the densities' precise forms. Instead of bounding $\langle W_{\rm diss} \rangle$, one can calculate a one-shot dissipated work exactly.

One might worry that 
the RHS of Eq.~\eqref{eq:KawaiOne1} diverges.
For instance, a point particle has a Dirac-delta-function $\rho$,
if the particle has a particular momentum.
Evaluating $D_\infty$ on a divergent $\rho$ would yield infinity.
In reality, however, finite precision limits measurements of a particle's position and momentum.
This practicality regulates the divergence,
 rendering Theorem~\ref{theorem:Kawai} applicable to realistic particles.

Interchanging the arguments of $D_\infty$ yields the worst-case forfeited work. One can extract less work by implementing the reverse protocol at finite speed than by implementing the protocol quasistatically, due to dissipation. The {\em worst-case forfeited work} 
\begin{align}   \label{eq:Forfeit}
   W_\forfeit^{\rm worst}  := \Delta F  -  W_{\rm max}
\end{align}
is the most work an agent might sacrifice for time in any finite-speed reverse trial: 
\begin{align}
   W_\forfeit^{\rm worst} \label{eq:KawaiOne2}
   = \frac{1}{\beta}
   D_\infty   \LParen\tilde{\rho} (q, -p, t) ||   \rho(q, p, t)\RParen,
	\end{align} 
if ${\rm supp} \LParen\tilde{\rho} (q, -p, t)\RParen\subseteq{\rm supp}\LParen\rho(q, p, t)\RParen$.

%
%
% Quantum section
%
% 
\subsection{Divergences between quantum states}

Parrondo \emph{et al.}\ have quantized Eq.~\eqref{eq:Kawai}~\cite{ParrondoVdBK09}. They consider a quantum system governed by a quantum Hamiltonian $H(\lambda_t)$ specified by an external parameter $\lambda_t$. Let $\rho(t)$ denote the state occupied by the system at time $t$. In the forward protocol, the system begins in thermal equilibrium: 
$\rho(-\tau)  =  e^{- \beta H_{-\tau} } / Z_{-\tau}$.
During $t \in (-\tau, \tau)$, the system is isolated from the bath, and an agent invests work to switch $\lambda_t$ from $\lambda_{-\tau}$ to $\lambda_\tau$. The state changes unitarily.
During the reverse protocol, the system is prepared in the state $\tilde{\rho}( \tau )  =  e^{- \beta H_\tau } / Z_\tau$; time runs from $t = \tau$ to $t = -\tau$; and work is extracted via the time-reversed schedule $\lambda_{-t}$. 

Assuming that ${\rm supp}\LParen\rho(t)\RParen\subseteq  {\rm supp}\LParen\tilde{\rho}(t)\RParen$,
Parrondo \emph{et al.}\ derive 
\begin{align}   \label{eq:PResult}
   \langle W_\diss \rangle
   =  \frac{1}{\beta} \Rel \big(  \,   \rho(t) || \tilde{\rho} (t)    \,  \big).
\end{align}
% Reference: Parrondo, p. 11, top of the page

Recycling their set-up, we will prove a proportionality between the worst-case dissipated work and an order-$\infty$ R\'{e}nyi divergence. 
We must define ``work'' explicitly. 
In some quantum fluctuation-relation contexts, work is defined in terms of two energy measurements~\cite{Tasaki00,Kurchan00}: The system begins in the thermal state $\gamma_{-\tau}$. An energy measurement at $t = -\tau$ yields some eigenvalue $E_i$ of $H_{-\tau}$. 
The system is isolated from the bath, and the state evolves unitarily. 
An energy measurement at $t = \tau$ yields some eigenvalue $\tilde{E}_j$ of $H_\tau$. 
As the system exchanges no heat during the unitary evolution, the difference between the measurement outcomes equals the work performed: $W = \tilde{E}_j - E_i$. 

We assume that the agent does not learn the initial measurement's outcome until the end of the protocol. Because the state begins block-diagonal relative to the initial Hamiltonian, this measure-and-forget operation preserves the initial state.

\begin{theorem}   \label{theorem:QResult}
The worst-case work dissipated during any such quantum forward trial is
\begin{equation}  \label{eq:QWorst}
   W_\diss^{\rm worst}   =
   \frac{1}{\beta}   D_\infty  \LParen{\rho}(t) \, || \,\tilde{\rho}(t)\RParen.
\end{equation} 

\begin{proof}
Let $\rho(t) = \sum_i p_i \ketbra{i(t)}{i(t)}$ and 
$\tilde{\rho}  =  \sum_j \tilde{p}_j \ketbra{\tilde{j}(t)}{\tilde{j}(t)}$
denote the states' eigenvalue decompositions.
The eigenvalues, and the inner products
$\langle  i(t)  |  \tilde{j}(t)  \rangle$, remain constant throughout the unitary evolution.
$D_\infty\LParen\rho(t)   ||   \tilde{\rho}(t)\RParen$ therefore remains constant. 
Without loss of generality, we can evaluate the definition [Eq.~(\ref{eq:DefDinftyQuantum})] at $t = \tau$:
\begin{equation}
   \label{eq:QDinfInt}
   D_\infty   \LParen\rho(t)   ||   \tilde{\rho}(t)\RParen
   =   \ln   \left(   \max_{i,j}   
   \left\{  \frac{p_i}{\tilde{p}_j} \, :  \,   \braket{i(\tau)}{\tilde{j}(\tau)}
   \neq0   \right\}   \right).
\end{equation}

Let $U$ denote the unitary that evolves the initial state to the final in the forward process:
$\rho(\tau)  =  U \rho( - \tau )  U^\dag$. We can express the inner product as
$\langle  i ( - \tau )  |  U^\dag  |  \tilde{j} (\tau)  \rangle$. The thermal natures of $\rho(-\tau)$ and 
$\tilde{\rho} (\tau)$ imply that $p_i  = e^{-\beta E_i} / Z_{-\tau}$ and 
$\tilde{p}_j = e^{-\beta \tilde{E}_j} / Z_{\tau}$. Since $Z_\tau / Z_{-\tau} = e^{- \beta \Delta F}$, Eq.~(\ref{eq:QDinfInt}) is equivalent to
\begin{align}   \label{eq:QHelp}
   D_\infty \LParen\rho(t)  |  \tilde{\rho}(t)\RParen
   & = \ln   \Big(   \max_{i,j}  \Big\{  
      e^{ \beta ( \tilde{E}_j   -   E_i   -   \Delta F ) }  \,  :
      \nonumber  \\
   & \qquad   
      \langle  i ( - \tau )  |  U^\dag  |  \tilde{j} (\tau)  \rangle  \neq  0
   \Big\}   \Big).
\end{align}
The work dissipated in some forward trial is proportional to the exponential's argument.
The forward protocol is unable to map $\ket{  i ( - \tau ) }$ to $\ket{ \tilde{j} (\tau) }$ if and only if
$\bra{ i ( - \tau )}  U^\dag    \ket{\tilde{j} (\tau)} = 0$, i.e., if and only if the condition in Eq~.\eqref{eq:QHelp} is violated.
Hence the worst-case work that can be dissipated during any forward trial is proportional to exponential's argument, maximized under the condition in Eq.~\eqref{eq:QHelp}.
 Rearranging Eq.~\eqref{eq:QHelp} yields Eq.~(\ref{eq:QWorst}).
\end{proof}
\end{theorem}

%
%
% Analysis of QM 1-shot results
%
%
The discussion of irreversibility, distinguishability, $t$-dependence, the quasistatic limit, and coarse-graining that characterizes the classical Theorem~\ref{theorem:Kawai} characterizes also the quantum Theorem~\ref{theorem:QResult}. 
$W_\diss^{\rm worst}$ is bounded 
when $H_{-\tau}$ and $H_\tau$ have bounded spectra. 
Bounded spectra characterize many realistic systems, 
including one-shot problems (e.g.,~\cite{HorodeckiO13}).

An unbounded example seemingly curtails the theorem's applicability:
 the classical harmonic oscillator (HO). 
Specifically, take a positively charged classical particle that moves in one dimension (the $x$-axis),
 in a potential well centered at $x = 0$.
Consider turning on and off an electric field.
In the worst case, \emph{prima facie,} the field pushes the particle
to the top of the well---infinitely high up, costing $W_\diss^\worst = \infty$.
However, an HO accurately models a realistic particle only near $x = 0$.
Farther away, a realistic potential likely flattens, or turns over into a deeper potential well, 
or becomes well-modeled by an infinitely hard wall, etc.
In real-world situations, therefore, $W_\diss^\worst$ is finite.\footnote{
% < f >
Even extreme settings lead to finite $W_\diss^\worst$ values.
Consider, as an example, a work protocol $\mathcal{P}$
that preserves the system's volume, $V$.
The greatest amount $W$ of work that could be performed
would turn the system into a volume-$V$ black hole.
Adding more energy would raise the black hole's mass, $M$.
The mass varies directly with the radius, $R$: $M \propto R$.
Hence adding more energy would violate
the protocol's finite-volume constraint
and so would not be work performable during $\mathcal{P}$.
% The amount of energy required 
% to turn the system into a black hole
% is expected to cap $W_\diss^\worst$,
% if the system has a fixed finite size.
% Reference: GChat with Ning Bao -- 3/3/18 -- around 9:50 AM
Though this example might appear contrived,
it has relevance to contemporary physics:
The intersection of general relativity 
and quantum thermodynamics,
especially together with high-energy physics,
forms a frontier being explored now. 
Initial steps in this direction include~\cite{Bartolotta_18_Jarzynski,Cirstoiu_17_Irreversibility,Opatrny_12_Black}
and many works inspired by the black-hole-information paradox.
We leave a detailed unification of these fields
with one-shot statistical mechanics as 
an opportunity for further study.}
% < /f >

Let us apply Theorem~\ref{theorem:QResult} to a sudden quench.
Quantum quenches' work distributions have been studied
 in the context of the transverse-field Ising model~\cite{Silva_08_Statistics,Gambassi_11_Statistics,Abeling_16_Quantum},
 trapped ions~\cite{DornerGCPV12},
 randomly quenched finite-dimensional systems~\cite{Lobejko_17_Work}, 
 Fermi gases~\cite{Sindona_14_Statistics}, 
 and semiclassical approximations~\cite{Garcia_16_Quantum}.
Consider a finite-dimensional quantum system $\QSys$, 
e.g., a set of $\Sites$ qubits (two-level systems). 
Let $H(\lambda_t)$ denote the Hamiltonian.
The parameter $\lambda_t$ is {\em quenched} (changed instantaneously)
 from $\lambda$ to $\tilde{\lambda}$ during the forward protocol
 and from $\tilde{\lambda}$ to $\lambda$ during the reverse protocol~\cite{KawaiPV07}.
$\QSys$ begins the forward protocol in the state
 $\rho(-\tau)=e^{-\beta H(\lambda)}/Z_{-\tau}$,
 wherein $H(\lambda) = \sum_j E_j \ketbra{E_j}{E_j}$.
$\QSys$ begins the reverse protocol in 
 $\tilde{\rho}(\tau)=e^{-\beta H(\tilde{\lambda})}/Z_\tau$,
 wherein $H(\tilde{\lambda}) = \sum_j  \tilde{E}_j  \ketbra{\tilde{E_j}}{\tilde{E_j}}$.
By the density operator's statistical interpretation,
 $\QSys$ can be regarded as starting each trial
 in an energy eigenstate 
 chosen according to a Gibbs distribution.
% NYH: Why I included the above sentence, which might sound obvious: Not all researchers in related subfields realize that, and PRL paper must be broadly accessible. Example of unaware researchers: condensed-matter physicists. I know from recent experience...
% AJPG: I believe it! I think CMP and QI physicists lock horns more than any other pairing, because we study similar systems with completely different languages....
In the worst case, $\QSys$ begins in 
the forward process in the lowest-energy eigenstate of $H(\lambda)$, $\ket{E_\Min}$,
which is the highest-energy eigenstate of $H(\tilde{\lambda})$,
$\ket{\tilde{E}_\Max}$.
The work dissipated is $W_\diss^\worst=\tilde{E}_\Max-E_\Min-\Delta F$.
Now, we calculate $D_\infty$.
The state's form has no time to change during the quench.
Hence $\rho(t)=\rho(0)$ and $\tilde{\rho}(t)=\rho(\tau)$ $\forall t$. Therefore,
$D_\infty\LParen\rho(t)||\tilde{\rho}(t)\RParen=
\log\left(\min_{j,k} \left\{\lambda\in\mathbb{R}:\frac{e^{-\beta E_j}}{Z_{-\tau}}\leq\lambda\frac{e^{-\beta\tilde{E}_k}}{Z_\tau}\right\}\right)=\log\left(e^{-\beta E_\Min+\beta\tilde{E}_\Max}\right)+\log(Z_{-\tau}/Z_\tau)$.
Equation~\eqref{eq:QWorst} is satisfied.

%
%
% W_diss  +  work distributions
%
%
\subsection{Divergences between work distributions}

We have related dissipated work to a divergence $D_\infty$ between phase-space densities and to a $D_\infty$ between quantum states. We now relate $W_\diss^{\rm worst}$ to a $D_\infty$ between distributions over possible values of work. 

The Kullback-Leiber divergence between $P_\mathrm{fwd}(W)$ and $P_\mathrm{rev}(-W)$ is proportional to the average dissipated work: 
\begin{equation}  \label{eq:RelS2}
   \frac{1}{\beta} D\LParen P_{\rm fwd} (W)  ||  P_{\rm rev} (-W)\RParen
   = \langle W \rangle_{\rm fwd} - \Delta F 
   =  \langle W_{\rm diss}  \rangle
\end{equation}
\cite{GomezPB,Wu_05_Phase}.
The first equality follows from the substitution from Crooks' Theorem [Eq.~(\ref{eq:Crooks})] for 
$P_{\rm fwd} (W)  /P_{\rm rev} (-W)$ in the definition of 
$ D\LParen P_{\rm fwd}(W)  ||  P_{\rm rev}(-W)\RParen$. 
We will derive a one-shot analog of Eq.~(\ref{eq:RelS2}).

%
% D_\infty (P_f || P_r)
%
\begin{theorem}   \label{theorem:DMaxFR}
The worst-case work that can be dissipated in any forward trial is proportional to the order-$\infty$ R\'{e}nyi divergence between $P_{\rm fwd} (W)$ and $P_{\rm rev}( -W )$:
\begin{equation}   \label{eq:DMaxFR}
   W_\diss^{\rm worst}
   = \frac{1}{\beta} D_\infty  \LParen P_{\rm fwd} (W)   ||   P_{\rm rev}( -W )\RParen,
\end{equation}
if the set of possible work-values is bounded.

\begin{proof}
By the definition of $ D_\infty$, 
% Reference: Datta, P. 3
\begin{align}   \label{eq:DMax}
   & D_\infty\LParen P_{\rm fwd} (W)   ||   P_{\rm rev}( -W )\RParen   \\
   & = \ln   \LParen{\rm min}   \left\{   \lambda \in \mathds{R}   : 
        P_{\rm fwd} (W)   \leq   \lambda   P_{\rm rev}( -W )  
        \; \forall \, W   \right\}\RParen.  \nonumber
\end{align}
Let us solve for the minimal $\lambda$-value $\lambda_{\rm min}$ that satisfies the inequality. First, we check that we can divide the inequality by $P_{\rm rev}(-W)$.
Crooks' Theorem implies that
$P_\fwd(W)=e^{\beta(W-\Delta F)}P_\rev(-W)$.
By assumption, $P_\fwd(W)$ and $P_\rev (-W)$ are nonzero only if $W$ is finite. Also, $\Delta F$ is finite. Hence Crooks' Theorem implies that $P_\rev(-W) = 0$ if and only if $P_\fwd(W) = 0$. 
In this case, the inequality becomes $0 \leq \lambda \cdot 0$, which is satisfied by any finite $\lambda$ and so does not determine $\lambda_{\rm min}$.
To solve for $\lambda_{\rm min}$, we can restrict our focus to $P_{\rm rev}(-W)  \neq  0$, then divide each side of the inequality in Eq.~\eqref{eq:DMax} by $P_{\rm rev}(-W)$:
\begin{equation}
   \lambda_{\rm min}   \geq   \frac{   P_{\rm fwd}(W)  }{  P_{\rm rev}(-W)   }
   \quad \forall \: W.
\end{equation}

Substituting into the RHS from Crooks' Theorem yields   
$\lambda_{\rm min}   \geq   e^{ \beta (W - \Delta F) }$.
The bound saturates when $W$ assumes its maximal value $W_{\rm max}$:
$\lambda_{\rm min}   =   e^{ \beta ( W_{\rm max}   -   \Delta F )  }
=  e^{\beta W_\diss^{\rm worst}}$.
Substituting into Eq.~(\ref{eq:DMax}) yields Eq.~(\ref{eq:DMaxFR}).

\end{proof}
\end{theorem}

%
% Analysis: work-distribution relation
%
Just as $\frac{1}{\beta}\Rel\LParen P_{\rm fwd}(W) || P_{\rm rev}(-W)\RParen$  equals the average, over many trials, of dissipated work, 
$\frac{1}{\beta} D_\infty\LParen P_{\rm fwd}(W) || P_{\rm rev}(-W)\RParen$ equals the most work that could be dissipated in any trial. An agent can calculate this dissipated work upon inferring $P_\fwd$ and $P_\rev$ from experimental or simulation statistics.

Theorem~\ref{theorem:DMaxFR} contains a R\'{e}nyi divergence between work distributions, rather than a $D_\infty$ between phase-space distributions 
or a $D_\infty$ between quantum states. 
Hence Theorem~\ref{theorem:DMaxFR} governs more protocols than Theorems~\ref{theorem:Kawai} and~\ref{theorem:QResult}, as it describes all protocols---quantum or classical, regardless of whether the system exchanges heat while work is performed---that obey Crooks' Theorem.

Interchanging the divergence's arguments yields the worst-case forfeited work [Eq.~(\ref{eq:Forfeit})]:
\begin{align}  \label{eq:DMaxRF}
   W_\forfeit^\worst =
   \frac{1}{\beta} D_\infty\LParen P_{\rm rev}(-W)  ||  P_{\rm fwd}(W)\RParen.
\end{align}

\section{Outlook}

We have developed one-shot analogs of three relationships between the average dissipated work $\langle W_\diss \rangle$ and the average R\'{e}nyi divergence $\Rel$. 
We related the worst-case dissipated work $W_\diss^{\rm worst}$ to 
an order-$\infty$ R\'{e}nyi divergence $D_\infty$ 
between classical phase-space distributions, between quantum states, and to a $D_\infty$ between work distributions.
The triptych of theorems demonstrates an unexpected generality of the equality
$W_\diss^\worst  =  \frac{1}{ \beta }  \:  D_\infty ( . || . )$.
% AJPG: Amazing word. I'm now envisioning our letter as painted by Hieronymus Bosch...
% NYH:  :D

Beyond this theoretical contribution, our results may 
have applications to experiments and simulations.
We applied Theorem~\ref{theorem:QResult} to a quantum quench,
 whose work distribution has been studied in diverse settings~\cite{Silva_08_Statistics,Gambassi_11_Statistics,Abeling_16_Quantum,DornerGCPV12,Lobejko_17_Work,Sindona_14_Statistics,Garcia_16_Quantum}.
%whose work distribution has been studied in several settings:
%the transverse-field Ising model~\cite{Silva_08_Statistics,Gambassi_11_Statistics,Abeling_16_Quantum},
%trapped ions~\cite{DornerGCPV12},
%randomly quenched finite-dimensional systems~\cite{Lobejko_17_Work}, 
%Fermi gases~\cite{Sindona_14_Statistics}, 
%and semiclassical approximations~\cite{Garcia_16_Quantum}.
Work distributions have been studied also for 
trapped ions~\cite{An_15_Experimental}, single-electron boxes~\cite{Saira_12_Test},
and classical gases~\cite{Crooks_07_Work}.
% Other applicable settings include classical-gas work distributions~\cite{Crooks_07_Work}, and experimental platforms such as} trapped ions~\cite{An_15_Experimental} and single-electron boxes~\cite{Saira_12_Test}.

%Also classical-gas work distributions have been calculated analytically~\cite{Crooks_07_Work}.
%Additionally, fluctuation relations have been realized 
%with experimental platforms including
%trapped ions~\cite{An_15_Experimental} and single-electron boxes~\cite{Saira_12_Test}.
% Our results may be applied also to bit reset (Landauer erasure) and Szil\'ard work extraction~\cite{YungerHalpernGDV14}.

Applications to such settings could assume many forms.
%For instance, experimentalists simulate their systems before performing experiments.
%From simulations, one might infer the right-hand side of Eq.~\eqref{eq:KawaiOne1} or of Eq.~\eqref{eq:QWorst}.
For instance, experimentalists simulating their systems, before performing experiments,
 might infer the right-hand side of Eq.~\eqref{eq:KawaiOne1} or of Eq.~\eqref{eq:QWorst}.
The $W_\diss^\worst$ estimate could inform the preparation of work resources (e.g., a sufficiently charged battery)
 sufficient to ensure that any implementation of the protocol succeeds.
Also, dissipated work may manifest as heat.
Equipment such as transistors can break if inundated with too much heat.
Such equipment may be strengthened to withstand $W_\diss^\worst$.
Additionally, high-precision measurements of small heat quantities are being developed
 (e.g.,~\cite{Pekola_15_Towards}).
Our results could provide a ``sanity check'' on whether new instruments are working properly.
If the measured heat exceeds $W_\diss^\worst$, the instrument is likely malfunctioning.

A few practicalities merit consideration in applications of our theorems.
Consider applying Theorem~\ref{theorem:DMaxFR} to experimental data.
One measures $W$ in each of several trials,
 and bins the outcomes to form a histogram.
Only finitely many trials can be performed, so
 $P_\fwd(W)$ and $P_\rev(-W)$ are estimated with finite precision~\cite{Pohorille_10_Good,Jarzynski_06_Rare,YH_16_Number}.
Some $W=W_0$ bin in the $P_\rev(-W)$ histogram might have height zero,
 though $P_\rev(-W_0)\neq0$.
The worst-case work would appear to diverge.
Given physical expectations that $W_\diss^\worst \neq \infty$, 
one could vary the histograms' bin widths
% NYH: You *could* vary the bin widths. Someone might come up with a better idea, though. The "could" allows for multiple alternatives.
% one would need to vary the histograms' bin widths
to model $P_\rev(-W)$ better.

% AJPG: I recommend omitting this part. It's good exposition, but more detail than necessary for this brief and word-constrained discussion:-
%
%For instance, a typical forward trial costs more work than a typical reverse trial yields~\cite{Jarzynski_06_Rare}.
%$P_\fwd(W)$ can have much weight on large $W$-values
% on which $P_\rev(-W)$ has little weight~\cite{YH_16_Number}.
%Let $W_\fwd^\Max$ and $W_\rev^\Max$ denote the greatest $W$-values
% associated with finite-height histogram bins.
%Likely, $W_\fwd^\Max > W_\rev^\Max$ but $P_\rev ( - W_\fwd^\Max ) \neq 0$.
%The reverse-process histogram's $W_\rev^\Max$ bin 
% may be extended rightward to $W_\fwd^\Max$.
%}

Relatedly, the histograms can be \emph{smoothed}.
An agent can trade off the guarantee that each trial will accomplish its purpose 
for the possibility of paying less work (or extracting more work). 
 %than by exerting caution. 
An agent's risk tolerance can be quantified with a parameter $\epsilon \in [0, 1]$. 
The agent ignores area-$\epsilon$ tails of the distributions,
 because they correspond to highly unlikely $W$-values~\cite{DahlstenCBGYHV17}.
This process, called \emph{smoothing}, has been introduced into R\'{e}nyi divergences~\cite{Datta09} 
 and into one-shot statistical mechanics (e.g.,~\cite{HorodeckiO13,Aberg13,Brandao_15_Second,Salek_15_Deterministic,vanderMeer_17_Smoothed}).
Smoothing offers a theoretical and practical opportunity 
 to advance this article's results further into applications.

%
%
% Note added
%
%
\textbf{Note added:}
Theorem~\ref{theorem:DMaxFR} appeared in a preprint of~\cite{YungerHalpernGDV14}, 
not in the published article. 
Since the first preprint of this article appeared,~\cite{Alhambra_16_Fluctuating,Aberg_16_Fully,DahlstenCBGYHV17,Wei_17_Relations,Guo_17_Demonstration} 
have addressed other aspects of the one-shot-and-fluctuation-relation
overlap. 
R\'enyi divergences were applied to
fluctuation relations within a resource-theory model in~\cite{Salek_15_Deterministic}.
\vspace{1.3em}

%
%
% Acknowledgements
%
%
% \acknowledgements
\begin{acknowledgments}
This work was supported by a Virginia Gilloon Fellowship; an IQIM Fellowship; a Barbara Groce Fellowship; a KITP Graduate Fellowship; NSF grants PHY-0803371, PHY-1125565, and PHY-1125915; the FQXi Large Grant for ``Time and the Structure of Quantum Theory''; the EPSRC; the John Templeton Foundation; the Leverhulme Trust; the Oxford Martin School; the NRF (Singapore); and the MoE (Singapore). 
The Institute for Quantum Information and Matter (IQIM) is an NSF Physics Frontiers Center with support from the Gordon and Betty Moore Foundation (GBMF-2644).
VV and OD acknowledge funding from the EU Collaborative Project TherMiQ (Grant Agreement 618074).
NYH thanks Ning Bao for conversations about high-energy scenarios.
We thank all our referees for feedback that enhanced this article.
\end{acknowledgments}

%
%
% Bibliography
%
%

\end{document}